# Imperceptible CMOS camera dazzle for adversarial attacks on deep neural networks

**Zvi Stein\* and Adrian Stern**

Ben Gurion University, /Electro-Optics Engineering Department, School of Electrical and Computer Engineering, Beer Sheva, Israel
\* Correspondence author.

**Abstract— Despite the outstanding performance of deep neural networks, they are vulnerable to adversarial attacks. While there are many invisible attacks in the digital domain, most physical world adversarial attacks are visible. Here we present an invisible optical adversarial attack that uses a light source to dazzle a CMOS camera with a rolling shutter. We present the photopic conditions required to keep the attacking light source completely invisible while sufficiently jamming the captured image so that a deep neural network applied to it is deceived.**

1. INTRODUCTION

Deep Neural Networks (DNNs) offer cutting-edge performance in image analysis and processing. However, they have been proven to be vulnerable to adversarial attacks [1]. These attacks involve introducing small perturbations into the input signal, causing the DNNs to deviate from the intended output. Adversarial examples, referred to as attacked images, were first demonstrated in DNN applications a decade ago by Szegedy et al. [2]. Since then, numerous approaches have been developed to discover various examples [3].

The possibility of finding adversarial examples is grounded in the assertion that networks do not learn the true features of an image. Instead, their classification is based on superficial characteristics, akin to what Goodfellow et al. described as a "Potemkin village" [4]. Accordingly, it is possible to discover two visually identical images with different classifications. This underscores the importance of keeping the perturbations minimal in size while ensuring that the two images, the perturbed and the original, are indistinguishable to the human eye.

Most adversarial attacks have been developed for the digital domain, that is, attacks applied to digital images. Yet, there is an ongoing effort to develop attacks for the physical domain as well [5] . For example, there have been demonstrated attacks in the physical world by temporarily illuminating a sign [6], or by cover the camera lens with a digital patch [7], adding an item to an object such as a sticker on a traffic sign, or wearing an object (glass or earrings, for example), or sticking LEDs [8] in the face area. There are examples of making changes to a three-dimensional element in the target to cause classifiers to detect it as a different object [9]. In this work, we develop a visually imperceptible attack that exploits the rolling shutter mechanism that is common with CMOS cameras.

The primary interest in the rolling shutter effect in CMOS cameras was to handle distortions obtained from photographing fast-moving objects moving at a frequency close to the scanning frequency [10]. Consequently, accurate models and simulations have been developed to allow the correction of the distortion. In [11] it was proposed to use a cellphone camera as a detector (receiver) for the communication of visual signals so that the temporal signal is converted to a spatial signal in the plane of the camera detector.

Adversarial attacks exploiting the rolling shutter have been presented in [12] and [13], considering the scenario shown in Fig. 1(a). The attack is performed by illuminating the target with modulated light so that the rolling shutter yields an image pattern that depends on the row exposed during the light pulses.  In [12]an attack was developed for the first time as a back door in the configuration of a black box, and later, [13] presents an attack configuration of a white-box attack, in the same setup of a light source that illuminates the scene that the camera sees. Even though the authors design the light pulses to be at high frequencies, such that flickering won't be observed, the ambient illumination is observable. A notable disadvantage in this setup is the requirement to illuminate the entire scene accurately.

Fig. 1(b) shows an attack scenario where an Attacking MOdulated Light Source (AMOLS) directly illuminates the camera, and due to the scanning of the rolling shutter, the light pulses appear as a pattern of stripes. Here, only the camera aperture is illuminated, and since the light is directed to the camera (rather than using a reflected light), much lower powers are required than with the setup in Fig. 1(a). Kohler et al. [14] showed such *at a camera* attack that exploits the rolling shutter mechanism; however, they did not attempt to make their source imperceptible.

In this article, we will explore the relationship between the human eye's ability to distinguish the interference transmitted from a light source *at the camera*, and the ability of a network to classify a target under the influence of the laser source.

A successful invisible attack occurs when the AMOLS beam covers the camera aperture (see Fig. 1 (b)), the peak irradiance is sufficient to temporarily dazzle the sensor, but the average irradiance is below the eye's sensitivity threshold.

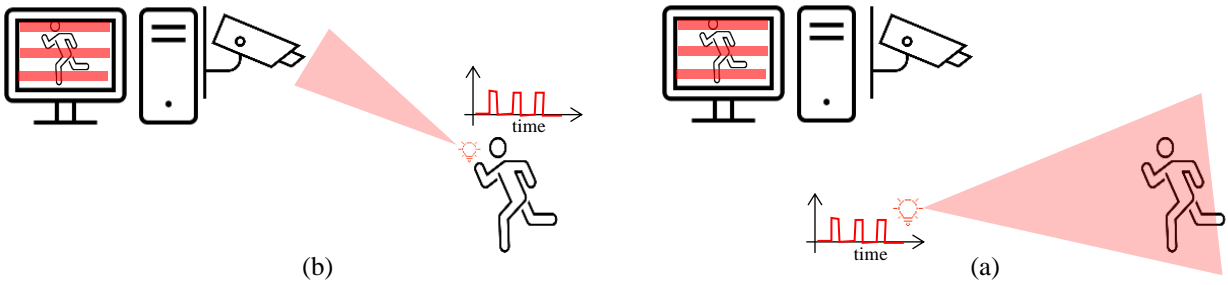

Fig. 1. Adversarial attack by temporally modulating the illumination of the entire scene (a), and by temporally dazzling the camera.

Since the integration time of the human eye is far longer than the acquisition time of each row in a rolling shutter camera, a high-frequency signal is seen as a continuous signal by the human eye. If the duty cycle of the AMOLS is $d.c.$, the intensity at the eye is roughly described by:

$$I_{eye} = (d.c.) * I_{source}, \qquad (1)$$

that is, the eye sees only the average power of the signal. Therefore, it is possible to control the intensity seen by the human eye by appropriately reducing the duty cycle of the AMOLS.

In this paper we explore the relationship between the effectiveness of a duty cycle *at the camera* attack and the distinctness of the attack. In the first step, we review the effect of an AMOLS on the camera, and we determine the irradiance required to provide the wanted phenomenon to disrupt a camera. In the second step, we evaluate the dazzling irradiance on the human eye. We determine conditions that define the eye's ability to see and identify the source. Finally, after the constraints on the required irradiance are known, we examine the efficiency of image distortion due to the set of pulses on a known classification network (ResNet50). We examine this with simulations and experiments.

## 2. DAZZLE EFFECT WITH ROLLING SHUTTER CAMERA

The spatial spread of a point source in the image plane is conventionally described by the diffraction Point Spared Function (PSF), which is given by the Fourier transform of the entrance pupil. However, particularly for bright power sources (e.g., a laser source), other effects such as stray light scattering and halo [15] may occur in addition to the diffraction PSF, which may be considerably larger than the PSF. The dazzling effect is demonstrated in Fig. 2. Figure 2 shows the dazzling effect we measured on a DELL-INSPIRON camera (0.92 Megapixel Diagonal viewing angle Camera 88 degrees) by an AMOLS power of $5mW$ and ~3.5mm spot diameter. As it can be seen, there is a significant dazzling phenomenon at this power.

Early research on infra-red imagers [16], [17] empirically found that the diameter of the saturated area in the image plane, $x_{sat}$, is proportional to the third power of the laser irradiance, $I_0^{1/3}$:

$$x_{sat} \propto \left(\frac{I_0}{I_{sat}}\right)^{\frac{1}{3}}, \qquad (2)$$

where $I_{sat}$ is the saturation level. In this paper, we use the results on visible light CMOS camera [18], [19] showing that for the camera used there, a minimum average irradiance of $50\frac{mW}{cm^2}$ during each row exposure is required, and at least $0.1\frac{mW}{cm^2}$ pick irradiance to achieve dazzling with shorter pulses. We found experimentally that similar conditions hold for the camera we used (Fig. 2).

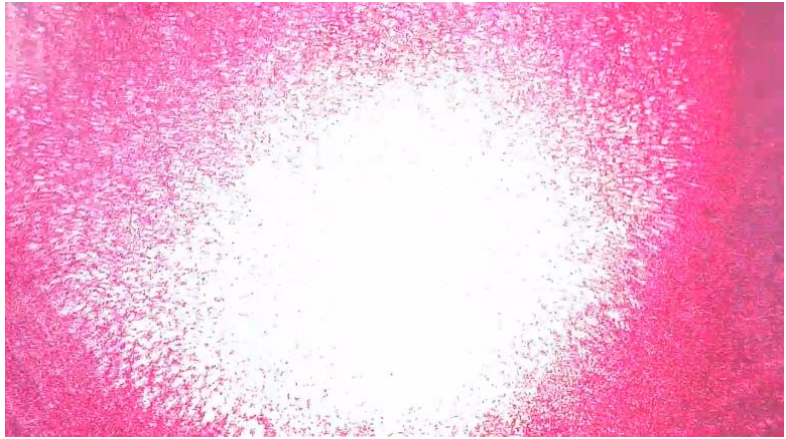

Fig. 2. Experimental measurement of the PSF phenomenon from a point source within the field of a camera, the flux measured in the object plane is ~$50\frac{mW}{cm^2}$.

Now let us see examine the dazzling effect as it appears in an attacked image. With a rolling shutter camera, every row in the frame collects the ambient light during a different time period. As illustrated in Fig. 3, the $i'th$ row of the sensor captures the light integrated during the period from $t_i - t_{exp}$ till $t_i$, while for the next row, $i + 1$, the integration time will be until $t_i + t_{read}$. We denote $t_{read}$ the reading time of a single row and $t_{exp}$ the exposure time of a single row, $N_r$ the number of rows in each frame, and $N_{rH}$ the number of hidden rows in each frame. The scanning duration $t_{frame}$ of scanning each frame is [10]:

$$t_{frame} = t_{read} * (N_r + N_{rH}) + t_{exp}. \qquad (3)$$

The number of exposed rows at any given time is given by the ratio $R_n = t_{exp}/t_{read}$ (seen Fig. 4). We will refer to $R_n$ as the rows exposure constant.

If the pulse generated by the AMOLS is shorter than $t_{read}$ then exactly $R_n$ rows are dazzled regardless of the pulse width. Therefore, for example, $1\mu s$ and $2\mu s$ pulses will produce the same pattern for a typical camera with $t_{read} \approx 30\mu s$.

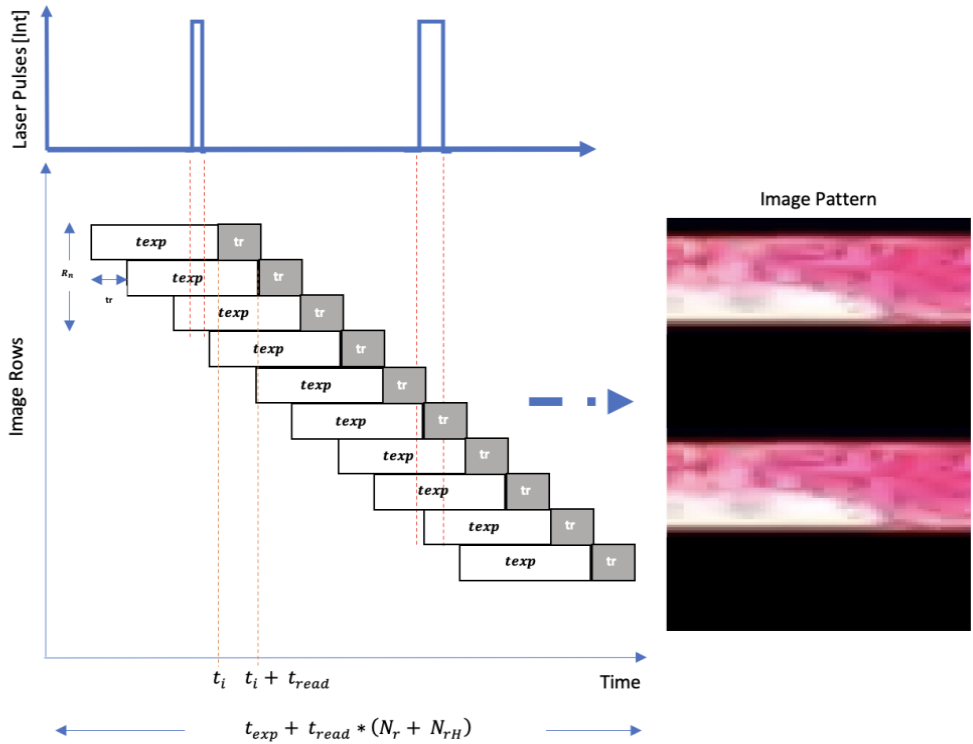

Fig. 3. A schematic description (left) of the relationship between the reading and exposure times and the line pattern obtained in the frame (right) under the dazzling of a modulated source (upper graph)

Fig. 4 demonstrates the dazzling pattern obtained on the rolling shutter sensor when an AMOLS is applied. The left-hand side in Fig. 4 shows the results of a real experiment, while the right-hand side shows the simulated pattern. For the simulations, we used $R_n = 37$ as obtained with a calibration process. The fit between simulated and real is 93%.

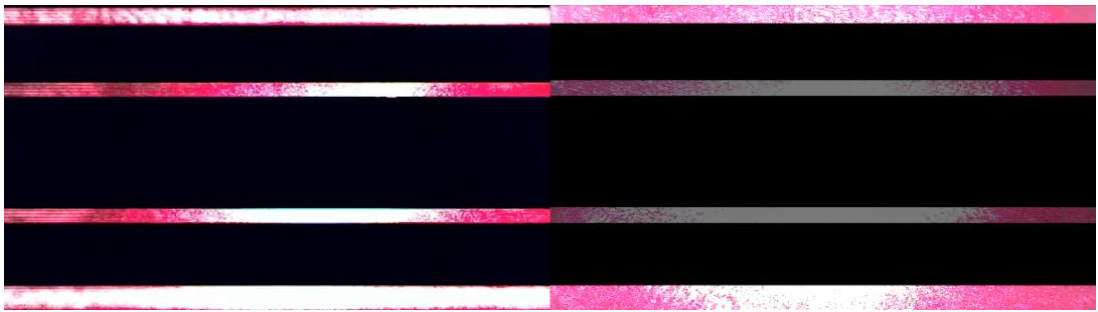

Fig. 4. A pattern of stripes obtained by simulation (right) and by our camera (left).

3. PHOTOPIC CONDITION FOR IMPERCEPTION

In this section, we shall determine the photometric conditions required to keep the AMOLS invisible. The geometry of the attacking scenario is shown in Fig. 5. This figure includes a target (car) and the AMOLS, in front of a camera that is supposed to classify the target. We suppose that the AMOLS power is set such that when it is on the irradiance at the sensor $e = 50 \frac{mW}{cm^2}$. A viewer (the eye in Fig. 5) is located near the camera and at an angle $\theta$ with respect of the optical axis. The average power $E$ from the AMOLS received by the eye depends on the duty cycle of the AMOLS over frame exposure. Furthermore, we assume that the AMOLS is smaller than the human eye's angular resolution. This is the strictest condition for the concentration of the seen power for any given source; an AMOLS covering a large angular extent yields a lower peak power.
The source duty cycle (d.c.) determines the minimum average power, $E = e * (d.c.)$.

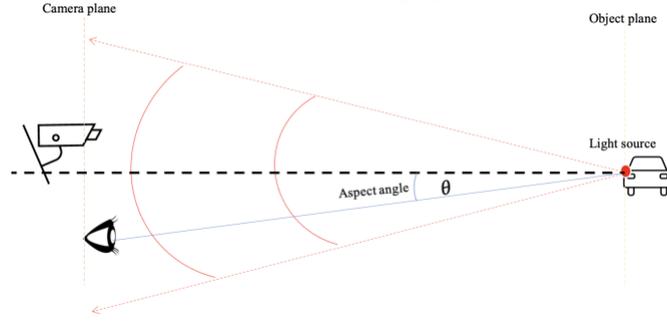

Fig. 5. A basic scenario in which the target (e.g., the car on the right) is located on the camera's (drawn on the left) optical axis, the purpose of the camera is to classify the target. The AMOLS attacks the camera while it should be invisible for a viewer at an angle θ with respect to the optical axis.

By denoting $L_b$ the background brightness and $L_{AS}$ the AMOLS brightness, the contrast is given by:

$$C_{orig} = \frac{L_{AS} - L_b}{L_b}. \quad (4)$$

Early work by Blackwell [20] and Werner [21] studied the threshold contrast $C_{thr}$ required to detect an object. According to Werner a small angle target contrast of 1 is sufficient to recognize the target.

Since the radiance is the physical unit that conserves along every optics system, its brightness is determined by it. One exception is when the solid angle covered by the target is less than the system's resolving power. In such a case, the background brightness is given [22]:

$$L_{AS} = E \cdot 683 \cdot V_\lambda \cdot \Omega_{eye}^{-2} \, [cd \cdot m^{-2}], \quad (5)$$

where $V_\lambda$ denotes the photopic efficacy and $\Omega_{eye}$ denotes the resolving power of the eye (which is the strictest assumption regarding the received power).
Employing a camera model for the eye model, Williamson and McLin [23], [24] suggested a scattering function based on empirical findings by Vos [25], with an effective solid angle that the eye collects:

$$f_{eye}(\theta, A, p, L_b) = S \cdot L_b^T \cdot g_{eye}(\theta, A, p) \, [sr^{-1}]. \quad (6)$$

The term $g_{eye}$ in Eq. (6) is determined by the off-axis angle $\theta$ (see Fig. 5) and age $A$ (in years), and the eye pigment $p$ is given by Eq. (7):

$$g_{eye}(\theta, A, p) = \frac{10}{\theta^3} + \left[\frac{5}{\theta^2} + \frac{0.1p}{\theta}\right]\left[1 + \left(\frac{A}{62.5}\right)^2\right] + 0.0025p. \, [sr^{-1}] \quad (7)$$

After substituting the term $E = e * (d.c.)$, and angular resolution by $f_{eye}$, Eq. (5) becomes:

$$L_{AS} = (D.C.) \cdot e \cdot 683 \cdot V_\lambda \cdot f_{eye} \, [cd \cdot m^{-2}]. \quad (8)$$

Finally, after substituting Eq. (8) in Eq. (4) we obtain:

$$d.c. = L_b^{1-T} \cdot \frac{C_{thr}(L_b) + 1}{e \cdot 683 \cdot V_\lambda \cdot S \cdot g_{eye}}. \quad (9)$$

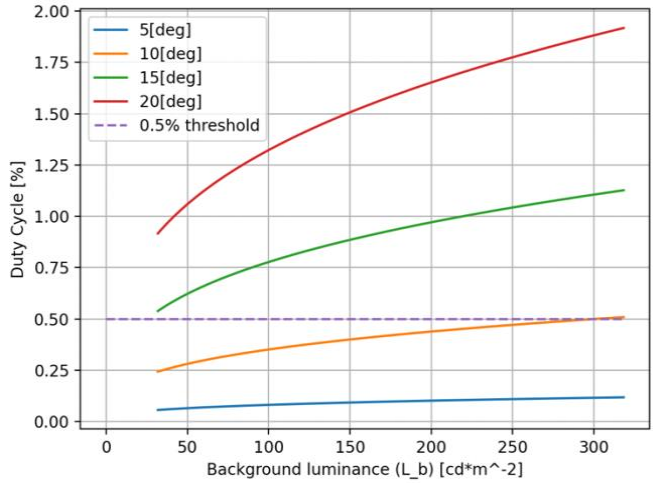

Fig. 6. The duty cycle of the AMOLS at the threshold of human discrimination, for various viweing angles θ.

Fig. 6 shows the d.c. required for dazzling as a function of the background illumination for various viewing aspect angles, α. The larger the aspect angle is, the effective radiance on the retina decreases. As the background brightness increases, the contrast decreases, and more power is required to pass the threshold. It can be seen in Fig. 6 that for viewers at α larger than 10 degrees, a duty cycle of 0.5% is sufficient for keeping the source invisible, regardless of the background illumination level. In the following Sections, we will see that we can provide a technique that works even at a lower percentage.

4. GENERATING THE ADVERSARIAL ATTACK

Following the formalism in [2] the problem of finding the adversarial example is:

$$\begin{aligned} &\text{Minimize } \left\lVert x' - x \right\rVert_2^2 \\ &\text{S.T. } C(x') \neq l \\ &x' \in [0, 1]^n, \end{aligned} \quad (10)$$

where $x$ denotes the undistributed image, $x'$ is the perturbated image, $l$ represents the ground truth label of image $x$, and the classifier is denoted by $C(x)$. Solving such a problem can be very complex, so it is necessary to convert it into a simpler problem suggested [26], described briefly in the following. Let $\delta = x' - x$ and $V_n$ denote the SoftMax of the logits of the network, and let us define its loss:

$$f(x) = Loss(V_n(x, l)). \quad (11)$$

The constraint minimization problem (10) can be replaced by:

$$\begin{aligned} &\text{Minimize } \left\lVert \delta \right\rVert_0 - \alpha \cdot f(x + \delta) \\ &\text{s.r. } x + \delta \in [0, 1]^n, \end{aligned} \quad (12)$$

where $\alpha$ defines the ratio between the magnitude of the disturbance and the intensity of the effect on the output, and $\lVert \cdot \rVert_0$ denotes the zero norm.

In our case, we aim to establish a relationship between laser activity and image perturbation. Let $\mathbf{E}_{eff}$ be an $N$ dimensional row binary vector determining the laser pulse activity, where $N = \frac{1}{R_n} \cdot (N_r + N_{rH})$, $R_n$ is the number of rows dazzled by each pulse, and $(N_r + N_{rH})$ is the total number of sensor rows (see Sec. 2). A unity value at the $i$'th component of this vector, $E_{eff}[i] = 1$, indicates a pulse at instant $t = i\frac{t_{frame}}{N}$. Such a pulse dazzles rows $(i \cdot R_n)$ to $(i+1) \cdot R_n$. Thus, the indexes of the image rows dazzled are given by replacing each unity entry in the laser pulse activity $\mathbf{E}_{eff}^T$ with $R_n$ ones, mathematically expressed by:

$$\mathbf{E}_r^T = \mathbf{E}_{eff}^T \otimes \mathbf{1}_{R_n}^T \quad (13)$$

where $\mathbf{1}_{R_n}^T$ is an $R_n$ - dimensional unity column vector and $\otimes$ is the Kronecker product. Hence, the resulting $\mathbf{E}_r^T$ is an $N$-dimensional binary column vector with unity values indicating the dazzled sensor's rows. The dazzled $N \times M$ image pattern (e.g., Fig. 4) can be obtained by

$$\delta = \mathbf{E}_r^T \otimes \mathbf{1}_M = \left(\mathbf{E}_{eff}^T \otimes \mathbf{1}_{R_n}^T\right) \otimes \mathbf{1}_M \quad (14)$$

where $\mathbf{1}_M$ is a unity row vector having the size of the image rows, $M$.

Now the minimization problem (12) can be expressed in terms of the laser pulse activity vector $\mathbf{E}_{eff}$:

$$\text{Minimize } \left\|\mathbf{E}_{eff}^T\right\|_0 - \alpha \cdot f(x + \delta)$$
$$\text{s.t. } \delta \in [0,1]^n. \quad (15)$$

In practice, to employ common gradient-based algorithms (such as SGD or ADAM) to solve (15), we replace the binary vector obtained (14) with:

$$\delta = \left[\left(\tfrac{1}{2}\tanh(\boldsymbol{\omega}^T) + 1\right) \otimes \mathbf{1}_{R_n}^T\right] \otimes \mathbf{1}_M \quad (16)$$

having values $\delta \in [0,1]^n$, and searching for the optimal solution in the $\boldsymbol{\omega}^T$ vector, which has the same dimensions as $\mathbf{E}_{eff}^T$.

Since the camera exposure instant is unknown to the attacker, the AMOLS pulse sequence generates a dazzling pattern with a random horizontal shift. To account for asynchrony between the attacking pulse sequence and the camera exposure instant, we employ the Expectation over Transformation (EoT) method [9], as follows:

$$\text{Minimize } \mathbb{E}_{t_0 \sim T}\left\{\left\|E_{eff}\right\|_0 - \alpha \cdot f(x + \delta)\right\} \quad (17)$$

where T is the space of all possible frame exposure instances, $t_0$.

## 5. RESULTS

In this section, we verify the attack performance obtained with the optimal image dazzle patterns found with the method in Section 4, respective to the AMOLS pulse activity defined in Sec. 2. We use simulation and real experiments. In Sec. 5.1 we use simulations to examine the role of the duty cycle while keeping the pulse width constant. In Sec. 5.2 we demonstrate the attack optically and study the sensitivity to the pulse width.

### 5.1 Simulation Study

We use experiments to examine the relationship between the effectiveness of the AMOLS attack and the duty cycle of the source (discussed in Sec. 4). The simulation for these experiments was conducted using a Python-based simulation, the specifics of which can be found in our publicly available code [27]. In our experiment, we use the ResNet50 classifier [28]. Figure 7 presents the standard cross-entropy loss as a function of the AMOLS duty cycle. The cross-entropy shown in the figure is the largest obtained for each duty cycle, as resulted from the optimization of the attack (Sec. 4). Fig. 7 presents simulation results for two cases; for an object that covers approximately 40% of the field of view (FOV) and for an object that covers approximately 85% of the FOV.

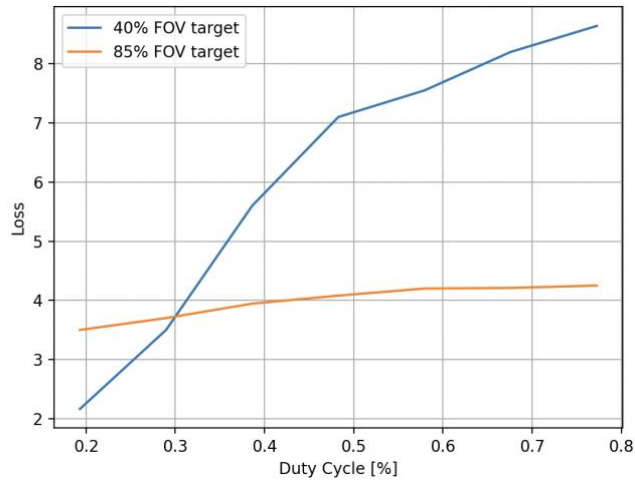

Fig. 7.    The effectiveness of the attack, the loss function as a function of the duty cycle.

The results shown in Fig. 7 are obtained for the "Coffee mug" label, for which we found empirically that with cross-entropy values larger than 2 the classifier begins to fail. It can be seen from the small object's (40% FOV) graph that a d.c. of 0.2% only marginally attacks the net, but by increasing the d.c. rapidly increases the loss deepening the attack. On the other hand, for larger objects (85% FOV) the attack is effective all over the presented range and the dependence on the d.c. is milder.

We have repeated simulated experiments for various objects, the objects in the images occupied between 15% - to 90% of the FOV. For each image, multiple attack realizations were applied. We found for 0.4% Duty-Cycle (equivalent to 4 pulses) the average attack success rate was found to be 90%.

*5.2 Real experiment results*

We placed a coffee mug in front of a camera DELL-INSPIRON camera (0.92 Megapixel Diagonal viewing angle Camera 88 degrees). Adjacent to the coffee mug, a laser diode (650nm Dot Diode Laser) with a power of 5mW and a spot size of 3.5mm was placed. The camera captured both the reflected light and the light received from the AMOLS. A sequence of pulses was designed to attack Resnet50 according to the optimization method in Sec. 4 and applied with the help of Arduino-Uno microcontroller board. The experiment was performed in the dark, with no ambient light, to provide ideal conditions for the classifier when the attack is not applied.

Fig. 8 demonstrates examples of the attacked images and the classification results. In this experiment, the AMOLS was activated at a d.c. of 0.01% and generated laser pulses of $1\mu s$. The attack for two different shots is shown. Recall that the attack is not synchronized to the camera; therefore, the attack appears differently in each image. As can be seen, the images are incorrectly classified, and the classification confidence level for the correct label (#500) is very low.

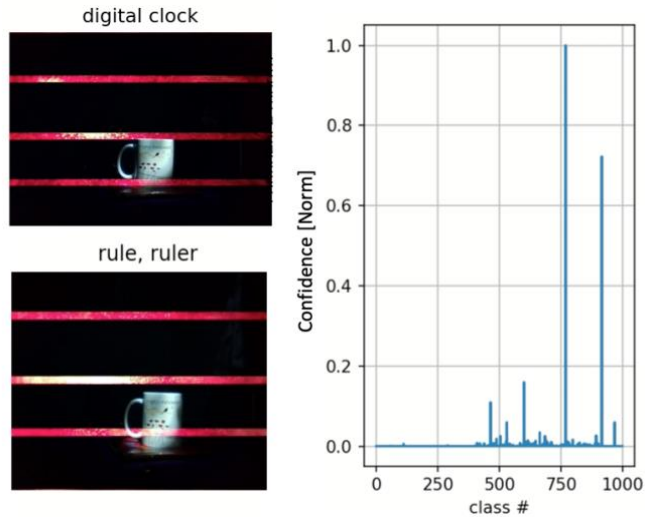

Fig. 8. Two attack examples output respective to two different shots (left), and the confidence level for the set of 1000 labels that the ResNet50 network was trained for (right). The correct label index of the "Coffee mug" is #500. AMOLS activity: pulse width of 1μs with d.c. of 0.01% (left above) and pulse width of 70μs with d.c. of 0.85% (left below).

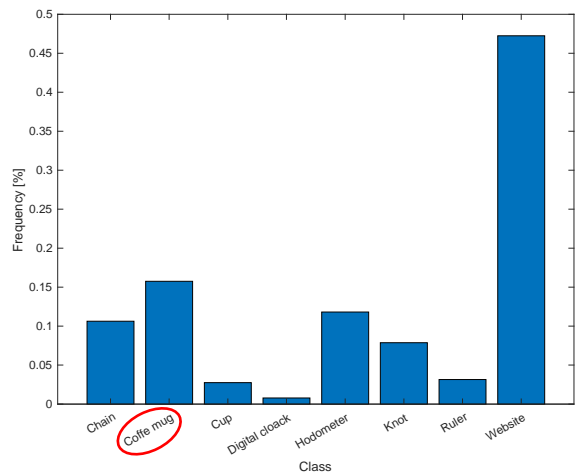

Fig. 9. The distribution of the predicted labels by the attacked classifier indicates the attack's success rate. The correct label for the object is "coffee mug," implying an attack success rate of 85%.

Fig. 9 shows the distribution of the predicted labels by the attacked classifier, as obtained from 254 repeated trails, using an attack employing 4 pulses with a width of $1\mu s$ The attack success rate in these conditions is 85%. An even higher success rate can be obtained by increasing the pulse width as shown in Fig. 10. As evident in Fig. 10, a pulse width larger than about $70\mu s$ an attack success rate of approaching 100%. However, we recall from Sec. 3, that increasing the pulse width reduces the concealed viewing angle $\theta$ (see Fig. 6). Let us evaluate this tradeoff for our experiment. Given our camera works at a frame rate of 30 frames per second, a pulse of $1\mu s$ accounts for a minimal d.c. of around $0.012\% = (100 \cdot 4 \cdot (1\mu s) \cdot (30s^{-1}))$, whereas $70\mu s$ constituted a higher d.c. of 0.85%. Examining Fig. 6, we see that for a d.c. of 0.01% the AMOLS's activity is not seen by viewers located within an angle larger than 5° with respect to the optical axis. For a laser fraction activity of 0.85% the imperceptive angel increases to 15°.

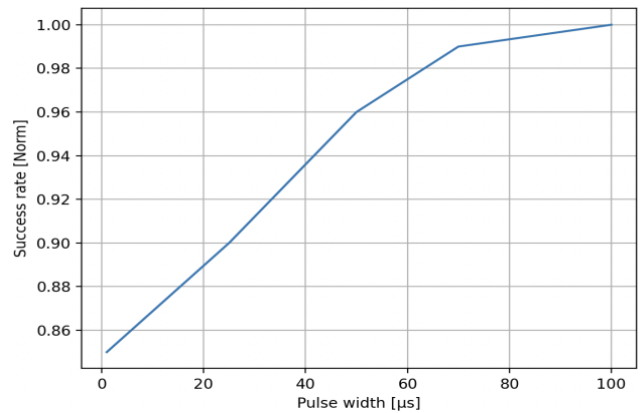

Fig. 10.    Average attack success rates as a function of the pulse width.

6. CONCLUSIONS

In conclusion, we demonstrated a DNN attack method for images captured with a rolling shutter camera. The attack is applied in a scenario where a light source sends a sequence of pulses toward the camera. The photometric and light pulse operations conditions that, on the one hand, dazzle the camera sufficiently to fool the DNN, and on the other hand, are determined to keep the light source's operation imperceptible.

We have shown that the light source d.c. (duty-cycle) controls the tradeoff between the attack's success and the light source's concealment. For example, with our system, we demonstrated an 85% attack success rate while keeping the light source activity imperceptible except for a small angular range of $5^0$ around the optical axis (Fig. 5). By sacrificing 10 degrees of concealment, the attack success rate can be increased to 98%.

7. DATA AND CODE AVAILABILITY STATEMENT

All data, simulation code, and video results supporting the findings of this study are openly available in our GitHub repository [27]. The repository includes all the materials necessary for replication and further exploration of the analyses presented in this paper.